\documentclass[10pt,letterpaper]{article}

\usepackage{cogsci}

\cogscifinalcopy % Uncomment this line for the final submission 

\usepackage{pslatex}
\usepackage{apacite}
\usepackage{float} % Roger Levy added this and changed figure/table
\usepackage{soul}
\usepackage{natbib}

\usepackage{amsmath}
\AtBeginDocument{%
  \setlength\abovedisplayskip{2pt}
  \setlength\belowdisplayskip{2pt}}

\usepackage{graphicx}
\usepackage{subcaption}
\usepackage{array}
\newcommand{\PreserveBackslash}[1]{\let\temp=\\#1\let\\=\temp}
\newcolumntype{C}[1]{>{\PreserveBackslash\centering}p{#1}}
\usepackage{dcolumn,booktabs}
\newcolumntype{d}[1]{D{.}{.}{#1}}
\newcommand\mc[1]{\multicolumn{1}{c}{#1}}

\usepackage[dvipsnames]{xcolor}

\title{Using Machine Learning to Guide Cognitive Modeling:\\A Case Study in Moral Reasoning}
 
\author{{\large \bf Mayank Agrawal (mayank.agrawal@princeton.edu)} \\
  Department of Psychology, Princeton University
  \AND {\large \bf Joshua C. Peterson (peterson.c.joshua@gmail.com)} \\
  Department of Computer Science, Princeton University
  \AND {\large \bf Thomas L. Griffiths (tomg@princeton.edu)} \\
  Departments of Psychology and Computer Science, Princeton University}

\begin{document}

\maketitle

\begin{abstract}

Large-scale behavioral datasets enable researchers to use complex machine learning algorithms to better predict human behavior, yet this increased predictive power does not always lead to a better understanding of the behavior in question. In this paper, we outline a data-driven, iterative procedure that allows cognitive scientists to use machine learning to generate models that are both interpretable and accurate. We demonstrate this method in the domain of moral decision-making, where standard experimental approaches often identify relevant principles that influence human judgments, but fail to generalize these findings to ``real world'' situations that place these principles in conflict. The recently released Moral Machine dataset allows us to build a powerful model that can predict the outcomes of these conflicts while remaining simple enough to explain the basis behind human decisions. 

%
% maybe one line for the 6-pager to save space. we can expand for 7-page final
\end{abstract}

\section{Introduction}

Explanatory and predictive power are hallmarks of any useful scientific theory. However, in practice, psychology tends to focus more on explanation \citep{yarkoni2017choosing}, whereas machine learning is almost exclusively aimed at prediction. The necessarily restrictive nature of laboratory experiments often leads psychologists to test competing hypotheses by running highly-controlled studies on tens or hundreds of subjects. Although this procedure gives a better understanding of the specific phenomenon, it can be difficult to generalize the findings and predict behavior in the ``real world,'' where multiple factors are interacting with one another. Conversely, machine learning takes full advantage of complex, nonlinear models that excel in tasks ranging from image classification \citep{krizhevsky2012imagenet} to video game playing \citep{mnih2015human}. The performance of these models scales with their level of expressiveness \citep{huang2018gpipe}, which results in millions of parameters that are difficult to interpret.

\begin{figure}[t]
\centering
    \includegraphics[trim={1.65cm 7.6cm 2.8cm 3.2cm},clip,width=0.7\linewidth]{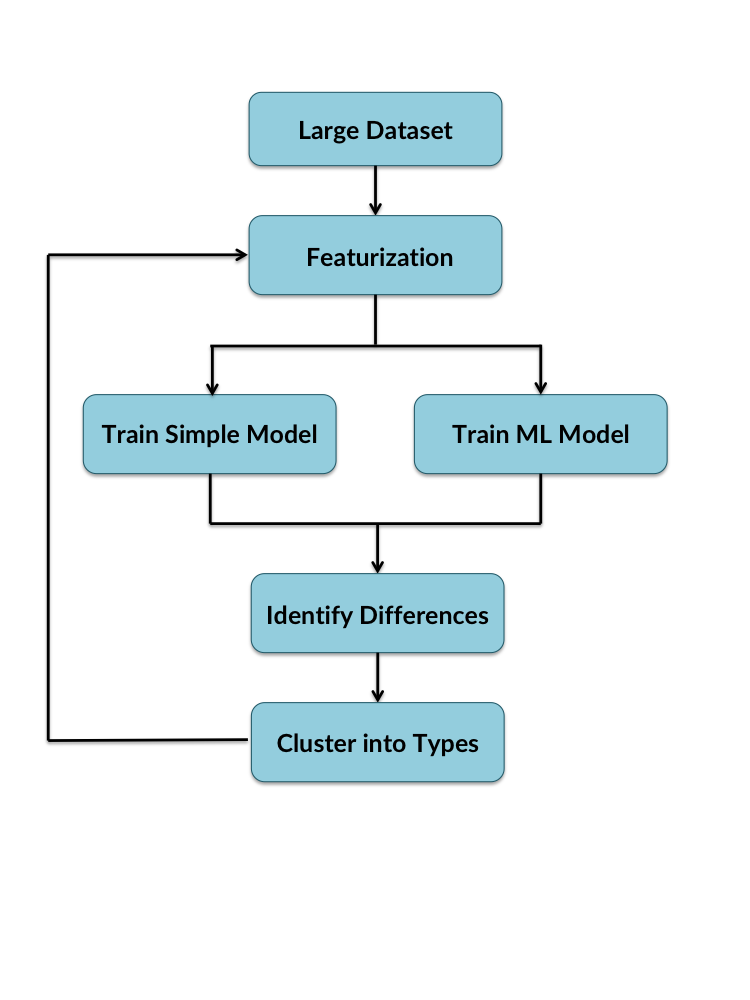}
    \caption{A systematic, data-driven procedure for building interpretable models that rival the predictive power of complex machine learning models.}
    \label{fig:flowchart}
    \vspace{-3mm}
\end{figure}

Interestingly, machine learning has long utilized insight from cognitive psychology and neuroscience \citep{rosenblatt1958perceptron,sutton1981toward,ackley1985learning,elman1990finding}, a trend that continues to this day \citep{banino2018vector,lazaro-gredilla_lin_guntupalli_george_2019}. We believe that the reverse direction has been underutilized, but could be just as fruitful. In particular, psychology could leverage machine learning to improve both the predictive and explanatory power of cognitive models. We propose a method (summarized in Figure \ref{fig:flowchart}) that enables cognitive scientists to use large-scale behavioral datasets to construct interpretable models that rival the performance of complex, black-box algorithms.

This methodology is inspired by Box's loop \citep{box1962useful,blei2014build,linderman2017using}, a systematic process of integrating the scientific method with exploratory data analysis. Our key insight is that training a black-box algorithm gives a sense of how much variance in a certain type of behavior can be predicted. This predictive power provides a standard for improvement in explicit cognitive models \citep{khajah2016deep}. By continuously critiquing an interpretable cognitive model with respect to these black-box algorithms, we can identify and incorporate new features until its performance converges, thereby jointly maximizing our two objectives of explanatory and predictive power.

In this paper, we demonstrate this methodology by building a statistical model of moral decision-making. Philosophers and psychologists have historically conducted thought experiments and laboratory studies isolating individual principles responsible for human moral judgment (e.g. consequentialist ones such as harm aversion or deontological ones such as not using others as a means to an end). However, it can be difficult to predict the outcomes of situations in which these principles conflict \citep{cushman2010our}. The recently released Moral Machine dataset \citep{awad2018moral} allows us to build a predictive model of how humans navigate these conflicts over a large problem space. We start with a basic rational choice model and iteratively add features until its accuracy rivals that of a neural network, resulting in a model that is both predictive and interpretable.

\section{Background}

\paragraph{Theories of Moral Decision-Making}

The two main families of moral philosophy often used to describe human behavior are \textit{consequentialism} and \textit{deontology}. Consequentialist theories posit that moral permissibility is evaluated solely with respect to the outcomes, and that one should choose the outcome with the highest value \citep{greene2007secret}. On the other hand, deontological theories evaluate moral permissibility with respect to actions and whether they correspond to specific rules or rights.

The trolley car dilemma \citep{foot_2002,thomson1984trolley} highlights how these two families differ when making moral judgments. Here, participants must determine whether it is morally permissible to sacrifice an innocent bystander in order to prevent a trolley car from killing five railway workers. The ``switch'' scenario gives the participant the option to redirect the car to a track with one railway worker, whereas the ``push'' scenario requires the participant to push a large man directly in front of the car to stop it, killing the large man in the process. Given that the outcomes are the same for the ``switch'' and ``push'' scenarios (i.e., intervening results in one death, while not intervening results in five deaths), consequentialism prescribes intervention in both scenarios. Deontological theories allow for intervening in the ``switch'' scenario but not the ``push'' scenario because pushing a man to his death violates a moral principle, but switching the direction of a train does not.

Empirical studies have found that people are much more willing to ``switch'' than to ``push'' \citep{greene2001fmri, cushman2006role}, suggesting deontological principles factor heavily in human moral decision-making. Yet, a deontological theory's lack of systematicity makes it difficult to evaluate as a model of moral judgment \citep{greene2017rat}. What are the rules that people invoke, and how do they interact with one another when in conflict? Furthermore, how do they interact with consequentialist concerns? Would people that refuse to push a man to his death to save five railway workers still make the same decision and with the same level of confidence when there are a million railway workers? Any theory of human moral cognition needs to be able to model how participants trade off different consequentialist and deontological factors.
  
\begin{figure}[t]
    \centering
    \begin{subfigure}[b]{0.4\textwidth}
        \includegraphics[width=\textwidth]{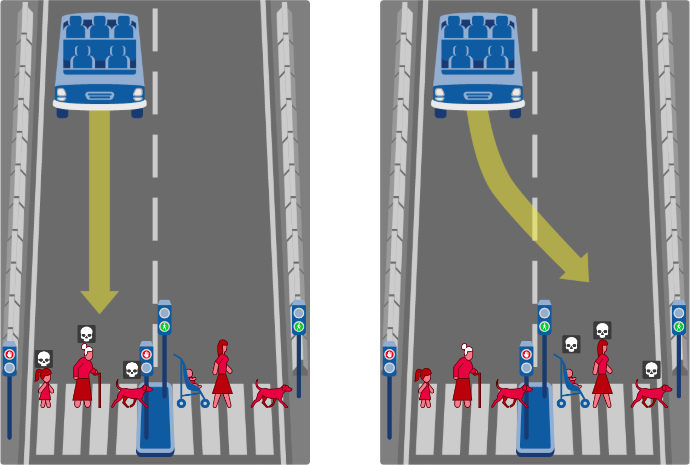}
        \caption{An autonomous car is headed towards a group of three pedestrians who are illegally crossing the street. The car can either stay and kill these pedestrians or swerve and kill three other pedestrians crossing legally.}
        \vspace{2mm}
        \label{fig:doubleped}
    \end{subfigure}
    ~ %add desired spacing between images, e. g. ~, \quad, \qquad, \hfill etc. 
      %(or a blank line to force the subfigure onto a new line)
    \begin{subfigure}[b]{0.4\textwidth}
        \includegraphics[width=\textwidth]{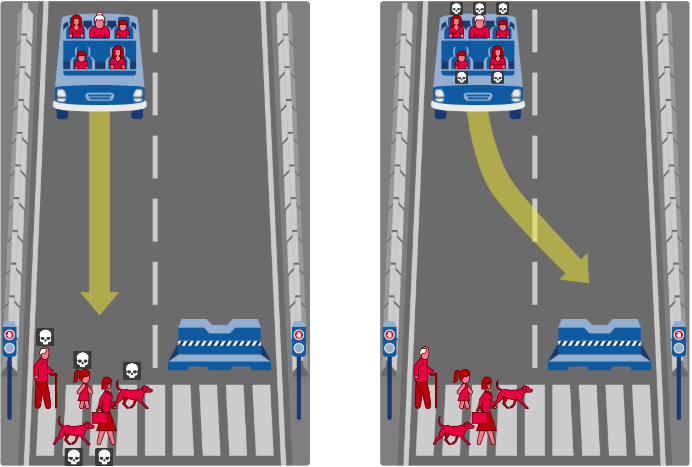}
        \caption{An autonomous car with five human passengers is headed towards a group of pedestrians who are illegally crossing the street. Staying on course will kill the pedestrians but save the passengers, while swerving will kill the passengers but save the pedestrians.}
        \label{fig:singleped}
    \end{subfigure}
    % ~ %add desired spacing between images, e. g. ~, \quad, \qquad, \hfill etc. 
    % %(or a blank line to force the subfigure onto a new line)
    \caption{Two sample dilemmas in the Moral Machine dataset. In every scenario, the participant is asked to choose whether to \textit{stay} or \textit{swerve} \protect\citep{awad2018moral}.}
    \label{fig:mmexample}
    \vspace{-6mm}
\end{figure}

\paragraph{Moral Machine Paradigm}
As society anticipates autonomous cars roaming its streets in the near future, the trolley car dilemma has left the moral philosophy classroom and entered into national policy conversations. A group of researchers aiming to gauge public opinion created ``Moral Machine,'' an online game that presents users with moral dilemmas (see Figure~\ref{fig:mmexample}) centered around autonomous cars \citep{awad2018moral}. Comprising roughly forty million decisions from users in over two hundred countries, the Moral Machine experiment is the largest public dataset collection on human moral judgment.

In addition to the large number of decisions, the experiment operated over a rich problem space. Twenty unique agent types (e.g. man, girl, dog) along with contextual information (e.g. crossing signals) enabled researchers to measure the outcomes of nine manipulations: action versus inaction, passengers versus pedestrians, males versus females, fat versus fit, low status versus high status, lawful versus unlawful, elderly versus young, more lives saved versus less, and humans versus pets. The coverage and density of this problem space provides the opportunity to build a model that predicts how humans make moral judgments when a variety of different principles are at play.

\section{Predicting Moral Decisions}

As described earlier, the iterative refinement method we propose begins with both an initial, interpretable model and a more predictive black-box algorithm. In this section, we do exactly this by contrasting rational choice models derived from moral philosophy with multilayer feedforward neural networks.

\subsection{Model Descriptions}

We restricted our analysis to a subset of the dataset ($N = 12,478,340$) where an empty autonomous vehicle must decide between saving the pedestrians on the left or right side of the road (see Figure~\ref{fig:doubleped} for an example). The models we consider below are tasked to predict the probability of choosing to save the left side.

\subsubsection{Interpretable Models} Choice models (CM) are ubiquitous in both psychology and economics, and they form the basis of our interpretable model in this paper \citep{luce1959individual,mcfadden1973conditional}. In particular, we assume that participants construct the values for both sides, i.e., $v_{\text{left}}$ and $v_{\text{right}}$, and choose to save the left side when $v_{\text{left}} > v_{\text{right}}$, and vice versa. The value of each side is determined by aggregating the utilities of all its agents:
\begin{equation} 
    \label{eq:value}
    v_{\text{side}} = \sum_i u_il_i
\end{equation}
where $u_i$ is the utility given to agent $i$ and $l_i$ is a binary indicator of agent $i$'s presence on the given side.

\cite{mcfadden1973conditional} proved that if individual variation around this aggregate utility follows a Weibull distribution, the probability that $v_{\text{left}}$ is optimal is consistent with the exponentiated Luce choice rule used in psychology, i.e.,
\begin{equation} 
    \label{eq:opt}
    P(v_{\text{left}} > v_{\text{right}}) = P(c = \text{left}|v_{\text{left}}, v_{\text{right}}) = \frac{e^{v_{\text{left}}}}{e^{v_{\text{left}}} + e^{ v_{\text{right}}}}
\end{equation}

In practice, we can implement this formalization by using logistic regression to infer the utility vector $\mathbf{u}$. We built three models, each of which provided top-down different constraints on the utility vector. Our first model, ``Equal Weight,'' required each agent to be equally weighted. At the other extreme, our ``Utilitarian'' model had no restriction. A third model, ``Animals vs. People,'' was a hybrid: all humans were were weighted equally and all animals were weighted equally, but humans and animals could be weighted differently.

Research in moral psychology and philosophy has found that humans use moral principles in addition to standard utilitarian reasoning when choosing between options \citep{quinn1989actions,spranca1991omission,mikhail2002aspects,royzman2002preference,baron2004omission,cushman2006role}. For example, one principle may be that allowing harm is more permissible than doing harm \citep{sep-doing-allowing}. In order to incorporate these principles, we moved beyond utilitarian-based choice models by expanding the definition of a side's value:
\begin{equation} 
    \label{eq:principles}
     v_{\text{side}} = \sum_i u_il_i + \sum_m \lambda_mf_m
\end{equation}
where $f_m$ is an indicator variable of whether principle $m$ is present on the side and $\lambda_m$ represents the importance of principle $m$. We built an ``Expanded'' model that introduces two principles potentially relevant in the Moral Machine dataset. The first is a preference for allowing harm over doing harm, thus penalizing sides that require the car to swerve in order to save them. Another potentially relevant principle is that it is more justified to punish unlawful pedestrians than lawful ones because they knowingly waived their rights when crossing illegally \citep{nino1983consensual}. This model was trained on the dataset to infer the values of $\mathbf{u}$ and $\mathbf{\lambda}$.

\subsubsection{Neural Networks} 
We use relatively expressive multilayer feedforward neural networks (NN) to provide an estimate of the level of performance that statistical models can achieve in this domain. These networks were given as inputs the forty-two variables that uniquely defined a dilemma to each participant: twenty for the characters on the left side, twenty for the characters on the right side, one for the side of the car, and one for the crossing signal status. These are the same inputs for the ``Expanded'' choice model. However, the ``Expanded'' model had the added restriction that the side did not change an agent's utility (e.g., a girl on the left side has the same utility as a girl on the right side), while the neural network had no such restriction.

The networks were trained to minimize the crossentropy between the model's output and human binary decisions. The final layer of the neural networks is similar to the choice model in that it is constructing the value of each side by weighting different features. However, in these networks, the principles are learned from the nonlinear interactions of multiple layers and the indicators are probabilistic rather than deterministic.

To find the optimal hyperparameters, we conducted a grid search, varying the number of hidden layers, the number of hidden neurons, and the batch size. All networks used the same ReLU activation function and and no dropout. Given that most of these models both performed similarly and showed a clear improvement over simple choice models, we did not conduct a more extensive hyperparameter search. A neural network with three 32-unit hidden layers was used for all the analyses in this paper.

\subsection{Model Comparisons}

\subsubsection{Standard Metrics} 

Table \ref{table:metrics} displays the results of the four rational choice models and the best performing neural network. All models were trained on eighty percent of the dataset, and the reported results reflect the performance on the held-out twenty percent. We report accuracy and area under the curve (AUC), two standard metrics for evaluating classification models. We also calculate the normalized Akaike information criterion (AIC), a metric for model comparison that integrates a model's predictive power and simplicity. All metrics resulted in the same expected ranking of models: Neural Network, Expanded, Utilitarian, Animals vs. People, Equal Weight.

\begin{table}[h]
\centering
\caption{Comparison of Standard Metrics}
\label{table:metrics}
\begin{tabular}{l *{4}{d{1.3}} }
\toprule
\bf{Model Type} & \mc{\bf{Accuracy}} & \mc{\bf{AUC}} & \mc{\bf{AIC}} \\ 
\midrule
Equal Weight  &  0.571    & 0.616  & 1.301  \\ %\cline{1-4}
Animals vs. People & 0.630 & 0.702 & 1.234 \\ %\cline{1-4}
Utilitarian   &  0.732     & 0.780 &  1.146 \\ %\cline{1-4}
Expanded  &   0.763      & 0.826 & 1.046  \\ %\cline{1-4}
Neural Network  &  0.774 & 0.845 & 0.983 \\ 
\bottomrule
\end{tabular}
\vspace{-2mm}
\end{table}

\subsubsection{Performance as a Function of Dataset Size}

\begin{figure*}[t]
\centering
\begin{subfigure}[b]{0.33\textwidth}
  \includegraphics[trim={3mm 4mm 1mm 4mm},clip,width=\textwidth]{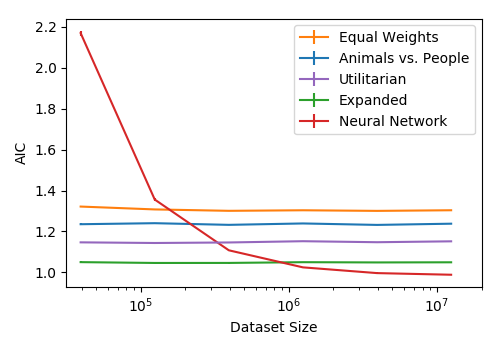}
  \caption{Dataset Size vs. AIC}
  \label{ex0}
\end{subfigure}\hfill
\begin{subfigure}[b]{0.33\textwidth}
  \includegraphics[trim={3mm 4mm 1mm 4mm},clip,width=\textwidth]{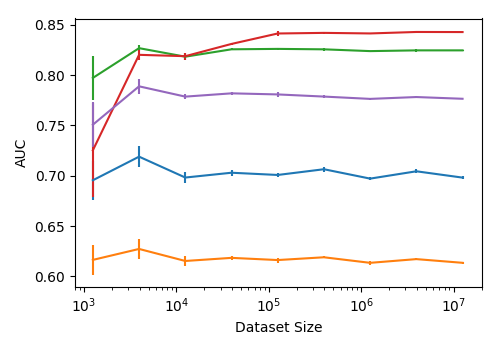}
  \caption{Dataset Size vs. AUC}
  \label{ex1}
\end{subfigure}\hfill
\begin{subfigure}[b]{0.33\textwidth}
  \includegraphics[trim={3mm 4mm 1mm 4mm},clip,width=\textwidth]{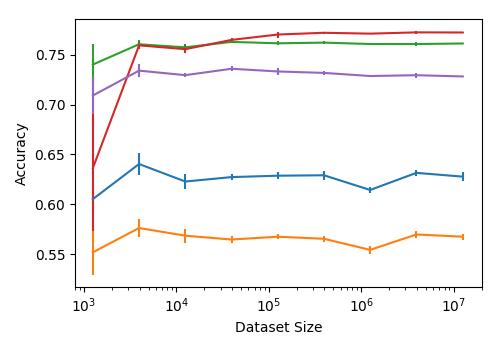}
  \caption{Dataset Size vs. Accuracy}
  \label{ex2}
\end{subfigure}
\vspace{-2mm}
\caption{Test-set performance metrics of choice models and neural network\footnotemark{} as a function of dataset size. Models were trained on five 80/20 training/test splits. Error bars indicate $\pm$1 SEM.}
\label{fig:metrics}
% \vspace{-2mm}
\end{figure*}

Table \ref{table:metrics} demonstrates that our cognitive models aren't as predictive as a powerful learning algorithm. This result, however, is only observable with larger datasets. Figure~\ref{fig:metrics} plots each metric for each model over a large range of dataset sizes. Choice models performed very well at dataset sizes comparable to that of a large laboratory experiment. Conversely, neural networks improved with larger dataset sizes until reaching an asymptote where $N > 100,000$, at which point they outperform rational choice models. These results suggest that while psychological models are robust in the face of small datasets, they need to be evaluated on much larger ones.

\section{Identifying Explanatory Principles}
The neural network gives us an aspirational standard of how our simpler model should perform. Next, our task is to identify the emergent features it constructs and incorporate them into our simple choice model.\\

% \paragraph{Calculating Residuals in Problem Aggregates}
\noindent\textbf{Calculating Residuals in Problem Aggregates}
By aggregating decisions for each dilemma, we can determine the empirical ``difficulty'' of each dilemma and whether our models predict this difficulty. For example, assume dilemmas A and B have been proposed to one hundred participants. If ninety participants exposed to dilemma A chose to save the left side and sixty participants exposed to dilemma B did, the empirical percentages for A and B would be $0.90$ and $0.60$, respectively. An accurate model of moral judgment should not only reflect the binary responses but also the confidence behind those responses.

We identified the specific problems where the neural network excelled compared to the ``Expanded'' rational choice model. Manually inspecting these problems and clustering them into groups revealed useful features beyond those employed in the choice model that the neural network is constructing. We formalized these features as principles and incorporated them into the choice model to improve prediction. Two examples are represented in Table~\ref{table:agg}.
\begin{table*}[t]
    \centering
    \caption{Problem Aggregate Comparisons (Left Side Save Percentage)}
    \vspace{-2mm}
    \label{table:agg}
    \begin{subtable}{\textwidth}
    \centering
    \begin{tabular}{p{5cm}p{4.4cm}p{1.4cm}C{1.4cm}C{1.4cm}C{1.4cm}}
    \toprule
    \textbf{Left Side Agents} & \textbf{Right Side Agents} & \textbf{Car Side} & \textbf{Empirical}    & \textbf{CM}  & \textbf{NN}       \\ 
    \midrule 
    Pregnant Woman Crossing Illegally & Cat Crossing Legally & Left & 0.779 & 0.411 & 0.797\\%932540
    Stroller Crossing Illegally & Cat Crossing Legally & Left & 0.826 & 0.425 & 0.801\\%790784
    Dog Crossing Legally & Male Doctor Crossing Illegally & Right & 0.312 & 0.693 & 0.293\\%1580699
    Cat Crossing Legally & Man Crossing Illegally & Right & 0.308 & 0.692 & 0.266\\%1566353
    Old Woman Crossing Illegally & Cat Crossing Legally & Left & 0.670 & 0.306 & 0.622\\%577072
    \bottomrule
    \end{tabular}
    \vspace{0.5mm}
    \caption{Problems indicating Human vs. Animals Principle}
    \label{tab:agghumansanimals}
    \end{subtable}% <---- don't forget this %
    \\
    \begin{subtable}{\textwidth}
    \centering
    \begin{tabular}{p{5cm}p{4.4cm}p{1.4cm}C{1.4cm}C{1.4cm}C{1.4cm}}
    \toprule
    \textbf{Left Side Agents} & \textbf{Right Side Agents} & \textbf{Car Side} & \textbf{Empirical}    & \textbf{CM}  & \textbf{NN}       \\ 
    \midrule 
    Old Man Crossing Legally & Boy Crossing Illegally & Right & 0.350 & 0.647 & 0.341\\%2146188
    Old Woman Crossing Legally & Girl Crossing Illegally & Right & 0.337 & 0.642 & 0.321\\%2063428
    Man & Boy & Left & 0.113 & 0.417 & 0.097\\%1311990
    Old Woman Crossing Legally & Girl Crossing Illegally & Left & 0.268 & 0.570 & 0.269\\%577077
    Old Woman & Woman & Right & 0.256 & 0.475 & 0.269\\%2067473
    \bottomrule
    \end{tabular}
    \vspace{0.5mm}
    \caption{Problems indicating Old vs. Young Principle}
    \label{tab:aggyoungold}
    \end{subtable}% <---- don't forget this %
\vspace{-6mm}
\end{table*}

Table \ref{tab:agghumansanimals} describes a set of scenarios where one human is crossing illegally and one pet is crossing legally. Empirically, users tend to overwhelmingly prefer saving the human, while the choice model predicts the opposite. Our choice model's inferred utilities and importance values reveal a strong penalty (i.e., a large negative coefficient) for (1) humans crossing illegally and (2) requiring the car to swerve. However, the empirical data suggests that these principles are outweighed by the fact that this is a humans-versus-animals dilemma, and that humans should be preferred despite the crossing or intervention status. Thus, the next iteration of our model should incorporate a binary variable signifying whether this is an explicit humans-versus-animals dilemma.

We can conduct a similar analysis for the set of scenarios in Table~\ref{tab:aggyoungold}. Both models output significantly different decision probabilities, the neural network being the more accurate of the two. Most salient to us was an effect of age. Specifically, when the principal difference between the two sides is age, both boys and girls should be saved at a much higher rate, and information about their crossing and intervention status is less relevant. To capture this fact, we can incorporate another binary variable signifying whether the only difference between the agents on each side is age.\\

% \paragraph{Incorporating New Features}
\noindent\textbf{Incorporating New Features}
The two features we identified are a subset of six ``problem types'' the Moral Machine researchers used in their experiment: humans versus animals, old versus young, more versus less, fat versus fit, male versus female, and high status versus low status. These types were not revealed to the participants, but the residuals we inspected suggest that participants were constructing them from the raw features and then factoring them into their decisions. 

\begin{figure*}[t]
\centering
\begin{subfigure}[b]{0.33\textwidth}
  \includegraphics[trim={3mm 4mm 1mm 4mm},clip,width=\textwidth]{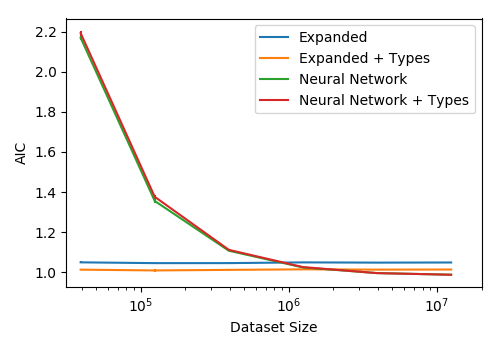}
  \caption{Dataset Size vs. AIC}
  \label{aictypes}
\end{subfigure}\hfill
\begin{subfigure}[b]{0.33\textwidth}
  \includegraphics[trim={3mm 4mm 1mm 4mm},clip,width=\textwidth]{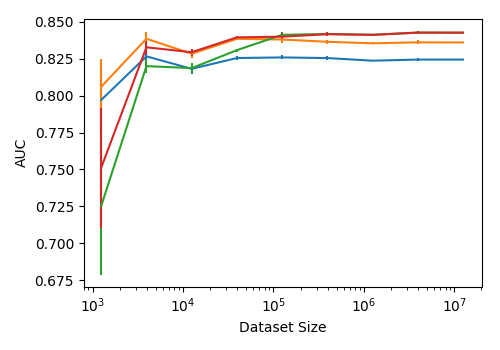}
  \caption{Dataset Size vs. AUC}
  \label{auctypes}
\end{subfigure}\hfill
\begin{subfigure}[b]{0.33\textwidth}
  \includegraphics[trim={3mm 4mm 1mm 4mm},clip,width=\textwidth]{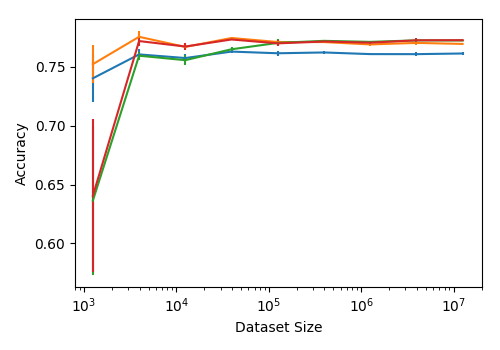}
  \caption{Dataset Size vs. Accuracy}
  \label{ex2}
\end{subfigure}
\vspace{-2mm}
\caption{Test-set performance metrics before and after incorporating new principles. Models were trained on five 80/20 training/test splits. Error bars indicate $\pm$1 SEM.}
\label{fig:metricsprinciples}
\vspace{-3mm}
\end{figure*}

Incorporating these six new features as principles resulted in 77.1\% accuracy, nearly closing the gap entirely between our choice model and neural network performance reported in Table \ref{table:metrics}. Figure \ref{fig:metricsprinciples} illustrates the effects of incorporating the problem types into both the choice model and the neural network in details.  Importantly, we observe that ``Neural Network + Types'' outperforms ``Neural Network'' at smaller dataset sizes, but performs identically at larger dataset sizes. This result suggests that the regular ``Neural Network'' is constructing the problem types we identified as emergent features given sufficient data to learn them from. More importantly, our augmented choice model now rivals the neural network's predictive power. And yet, by virtue of it being a rational choice model with only a few more parameters than our ``Expanded'' (and even the ``Utilitarian'') model, it remains conceptually simple. Thus, we have arrived at an interpretable statistical model that can both quantify the effects of utilitarian calculations and moral principles and predict human moral judgment over a large problem space.

Figure \ref{auctypes} still displays a gap between the AUC curves, suggesting there is more to be gained by repeating the process and potentially identifying new even more principles. For example, the last iteration found that when there was a humans-versus-animals problem, humans should be strongly favored. However, residuals suggest that participants don't honor this principle when all the humans are criminals. Rather, in these cases, participants may favor the animals or prefer the criminal by only a small margin. Thus, our next iteration will include a feature corresponding to whether all the humans are criminals. Our model also underperforms by overweighting the effects of intervention. In problem types such as male versus female and fat versus fit, the intervention variable is weighted much differently than in young-versus-old dilemmas. The next iteration of the model should also include this interaction. Thus, this methodology allows us to continuously build on top of the new features we identify.

\section{Conclusion}

\footnotetext{While a batch size of $8,192$ was used for Table \ref{table:metrics}, a batch size of $512$ was used here because of the smaller dataset sizes.}

Large-scale behavioral datasets have the potential to revolutionize cognitive science \citep{griffiths2015manifesto}, and while data science approaches have traditionally used them to predict behavior, they can additionally help cognitive scientists construct explanations of the given behavior. 

Black-box machine learning algorithms give us a sense of the predictive capabilities of our scientific theories, and we outline a methodology that uses them to help cognitive models reach these capabilities:
\begin{enumerate}
  \vspace{-1mm}
  \setlength\itemsep{-0.1em}
  \item Amass a large-scale behavioral dataset that encompasses a large problem space
  \item Formalize interpretable theories into parameterizable psychological models whose predictions can be evaluated
  \item Compare these models to more accurate, but less interpretable black-box models (e.g., deep neural networks, random forests, etc.)
  \item Identify types of problems where the black-box models outperform the simpler models
  \item Formalize these problem types into features and incorporate them into both the simple and complex models
  \item Return to Step 4 and repeat
  %
%   \vspace{-2mm}
  %
\end{enumerate}

We applied this procedure to moral decision-making, starting off with a rational choice model and iteratively adding principles until it had a comparable predictive power with black-box algorithms. This model allowed us to quantitatively predict the interactions between different utilitarian concerns and moral principles. Furthermore, our results regarding problem types suggest that moral judgment can be better predicted by incorporating alignable differences in similarity judgments \citep{tversky1993context}, such as whether the dilemma is humans-versus-animals or old-versus-young.

The present case study, while successful, is only a limited application of the methodology we espouse, and further demonstrations are required to illustrate its utility. It will be particularly interesting to apply our method to problems with even larger gaps between classic theories and data-driven predictive models. It is also likely that transferring insights from data-driven models will require moving beyond the sorts of featurization we consider here (i.e., problem clustering). In any case, we hope the microcosm presented here will inspire similarly synergistic approaches in other areas of psychology.\\

\vspace{-3mm}
\begin{small}
\noindent {\bf Acknowledgments.}
We thank Edmond Awad for providing guidance on navigating the Moral Machine dataset.
\end{small}
\vspace{-1mm}

% \clearpage
\bibliographystyle{apacite}

\setlength{\bibleftmargin}{.125in}
\setlength{\bibindent}{-\bibleftmargin}

% \clearpage
\renewcommand*{\bibfont}{\footnotesize}
\bibliography{references}

\begin{thebibliography}{}

\bibitem [\protect \citeauthoryear {%
Ackley%
, Hinton%
\BCBL {}\ \BBA {} Sejnowski%
}{%
Ackley%
\ \protect \BOthers {.}}{%
{\protect \APACyear {1985}}%
}]{%
ackley1985learning}
\APACinsertmetastar {%
ackley1985learning}%
\begin{APACrefauthors}%
Ackley, D\BPBI H.%
, Hinton, G\BPBI E.%
\BCBL {}\ \BBA {} Sejnowski, T\BPBI J.%
\end{APACrefauthors}%
\unskip\
\newblock
\APACrefYearMonthDay{1985}{}{}.
\newblock
{\BBOQ}\APACrefatitle {A learning algorithm for Boltzmann machines} {A learning
  algorithm for boltzmann machines}.{\BBCQ}
\newblock
\APACjournalVolNumPages{Cognitive science}{9}{1}{147--169}.
\PrintBackRefs{\CurrentBib}

\bibitem [\protect \citeauthoryear {%
Awad%
\ \protect \BOthers {.}}{%
Awad%
\ \protect \BOthers {.}}{%
{\protect \APACyear {2018}}%
}]{%
awad2018moral}
\APACinsertmetastar {%
awad2018moral}%
\begin{APACrefauthors}%
Awad, E.%
, Dsouza, S.%
, Kim, R.%
, Schulz, J.%
, Henrich, J.%
, Shariff, A.%
\BDBL {}Rahwan, I.%
\end{APACrefauthors}%
\unskip\
\newblock
\APACrefYearMonthDay{2018}{}{}.
\newblock
{\BBOQ}\APACrefatitle {The moral machine experiment} {The moral machine
  experiment}.{\BBCQ}
\newblock
\APACjournalVolNumPages{Nature}{563}{7729}{59}.
\PrintBackRefs{\CurrentBib}

\bibitem [\protect \citeauthoryear {%
Banino%
\ \protect \BOthers {.}}{%
Banino%
\ \protect \BOthers {.}}{%
{\protect \APACyear {2018}}%
}]{%
banino2018vector}
\APACinsertmetastar {%
banino2018vector}%
\begin{APACrefauthors}%
Banino, A.%
, Barry, C.%
, Uria, B.%
, Blundell, C.%
, Lillicrap, T.%
, Mirowski, P.%
\BDBL {}others%
\end{APACrefauthors}%
\unskip\
\newblock
\APACrefYearMonthDay{2018}{}{}.
\newblock
{\BBOQ}\APACrefatitle {Vector-based navigation using grid-like representations
  in artificial agents} {Vector-based navigation using grid-like
  representations in artificial agents}.{\BBCQ}
\newblock
\APACjournalVolNumPages{Nature}{557}{7705}{429}.
\PrintBackRefs{\CurrentBib}

\bibitem [\protect \citeauthoryear {%
Baron%
\ \BBA {} Ritov%
}{%
Baron%
\ \BBA {} Ritov%
}{%
{\protect \APACyear {2004}}%
}]{%
baron2004omission}
\APACinsertmetastar {%
baron2004omission}%
\begin{APACrefauthors}%
Baron, J.%
\BCBT {}\ \BBA {} Ritov, I.%
\end{APACrefauthors}%
\unskip\
\newblock
\APACrefYearMonthDay{2004}{}{}.
\newblock
{\BBOQ}\APACrefatitle {Omission bias, individual differences, and normality}
  {Omission bias, individual differences, and normality}.{\BBCQ}
\newblock
\APACjournalVolNumPages{Organizational Behavior and Human Decision
  Processes}{94}{2}{74--85}.
\PrintBackRefs{\CurrentBib}

\bibitem [\protect \citeauthoryear {%
Blei%
}{%
Blei%
}{%
{\protect \APACyear {2014}}%
}]{%
blei2014build}
\APACinsertmetastar {%
blei2014build}%
\begin{APACrefauthors}%
Blei, D\BPBI M.%
\end{APACrefauthors}%
\unskip\
\newblock
\APACrefYearMonthDay{2014}{}{}.
\newblock
{\BBOQ}\APACrefatitle {Build, compute, critique, repeat: Data analysis with
  latent variable models} {Build, compute, critique, repeat: Data analysis with
  latent variable models}.{\BBCQ}
\newblock
\APACjournalVolNumPages{Annual Review of Statistics and Its
  Application}{1}{}{203--232}.
\PrintBackRefs{\CurrentBib}

\bibitem [\protect \citeauthoryear {%
Box%
\ \BBA {} Hunter%
}{%
Box%
\ \BBA {} Hunter%
}{%
{\protect \APACyear {1962}}%
}]{%
box1962useful}
\APACinsertmetastar {%
box1962useful}%
\begin{APACrefauthors}%
Box, G\BPBI E.%
\BCBT {}\ \BBA {} Hunter, W\BPBI G.%
\end{APACrefauthors}%
\unskip\
\newblock
\APACrefYearMonthDay{1962}{}{}.
\newblock
{\BBOQ}\APACrefatitle {A useful method for model-building} {A useful method for
  model-building}.{\BBCQ}
\newblock
\APACjournalVolNumPages{Technometrics}{4}{3}{301--318}.
\PrintBackRefs{\CurrentBib}

\bibitem [\protect \citeauthoryear {%
Cushman%
, Young%
\BCBL {}\ \BBA {} Greene%
}{%
Cushman%
\ \protect \BOthers {.}}{%
{\protect \APACyear {2010}}%
}]{%
cushman2010our}
\APACinsertmetastar {%
cushman2010our}%
\begin{APACrefauthors}%
Cushman, F.%
, Young, L.%
\BCBL {}\ \BBA {} Greene, J\BPBI D.%
\end{APACrefauthors}%
\unskip\
\newblock
\APACrefYearMonthDay{2010}{}{}.
\newblock
{\BBOQ}\APACrefatitle {Our multi-system moral psychology: Towards a consensus
  view} {Our multi-system moral psychology: Towards a consensus view}.{\BBCQ}
\newblock
\APACjournalVolNumPages{The Oxford handbook of moral psychology}{}{}{47--71}.
\PrintBackRefs{\CurrentBib}

\bibitem [\protect \citeauthoryear {%
Cushman%
, Young%
\BCBL {}\ \BBA {} Hauser%
}{%
Cushman%
\ \protect \BOthers {.}}{%
{\protect \APACyear {2006}}%
}]{%
cushman2006role}
\APACinsertmetastar {%
cushman2006role}%
\begin{APACrefauthors}%
Cushman, F.%
, Young, L.%
\BCBL {}\ \BBA {} Hauser, M.%
\end{APACrefauthors}%
\unskip\
\newblock
\APACrefYearMonthDay{2006}{}{}.
\newblock
{\BBOQ}\APACrefatitle {The role of conscious reasoning and intuition in moral
  judgment: Testing three principles of harm} {The role of conscious reasoning
  and intuition in moral judgment: Testing three principles of harm}.{\BBCQ}
\newblock
\APACjournalVolNumPages{Psychological science}{17}{12}{1082--1089}.
\PrintBackRefs{\CurrentBib}

\bibitem [\protect \citeauthoryear {%
Elman%
}{%
Elman%
}{%
{\protect \APACyear {1990}}%
}]{%
elman1990finding}
\APACinsertmetastar {%
elman1990finding}%
\begin{APACrefauthors}%
Elman, J\BPBI L.%
\end{APACrefauthors}%
\unskip\
\newblock
\APACrefYearMonthDay{1990}{}{}.
\newblock
{\BBOQ}\APACrefatitle {Finding structure in time} {Finding structure in
  time}.{\BBCQ}
\newblock
\APACjournalVolNumPages{Cognitive science}{14}{2}{179--211}.
\PrintBackRefs{\CurrentBib}

\bibitem [\protect \citeauthoryear {%
Foot%
}{%
Foot%
}{%
{\protect \APACyear {2002}}%
}]{%
foot_2002}
\APACinsertmetastar {%
foot_2002}%
\begin{APACrefauthors}%
Foot, P.%
\end{APACrefauthors}%
\unskip\
\newblock
\APACrefYearMonthDay{2002}{}{}.
\newblock
{\BBOQ}\APACrefatitle {The Problem of Abortion and the Doctrine of the Double
  Effect} {The problem of abortion and the doctrine of the double
  effect}.{\BBCQ}
\newblock
\APACjournalVolNumPages{Virtues and Vices and Other Essays in Moral
  Philosophy}{}{}{19–32}.
\PrintBackRefs{\CurrentBib}

\bibitem [\protect \citeauthoryear {%
Greene%
}{%
Greene%
}{%
{\protect \APACyear {2007}}%
}]{%
greene2007secret}
\APACinsertmetastar {%
greene2007secret}%
\begin{APACrefauthors}%
Greene, J\BPBI D.%
\end{APACrefauthors}%
\unskip\
\newblock
\APACrefYearMonthDay{2007}{}{}.
\newblock
{\BBOQ}\APACrefatitle {The Secret Joke of Kant's Soul} {The secret joke of
  kant's soul}.{\BBCQ}
\newblock
\BIn{} W.~Sinnott-Armstrong\ (\BED), \APACrefbtitle {Moral Psychology: The
  Neuroscience of Morality: Emotion, Brain Disorders, and Development} {Moral
  psychology: The neuroscience of morality: Emotion, brain disorders, and
  development}\ (\BVOL~3, \BCHAP~2).
\newblock
\APACaddressPublisher{}{MIT Press}.
\PrintBackRefs{\CurrentBib}

\bibitem [\protect \citeauthoryear {%
Greene%
}{%
Greene%
}{%
{\protect \APACyear {2017}}%
}]{%
greene2017rat}
\APACinsertmetastar {%
greene2017rat}%
\begin{APACrefauthors}%
Greene, J\BPBI D.%
\end{APACrefauthors}%
\unskip\
\newblock
\APACrefYearMonthDay{2017}{}{}.
\newblock
{\BBOQ}\APACrefatitle {The rat-a-gorical imperative: Moral intuition and the
  limits of affective learning} {The rat-a-gorical imperative: Moral intuition
  and the limits of affective learning}.{\BBCQ}
\newblock
\APACjournalVolNumPages{Cognition}{167}{}{66--77}.
\PrintBackRefs{\CurrentBib}

\bibitem [\protect \citeauthoryear {%
Greene%
, Sommerville%
, Nystrom%
, Darley%
\BCBL {}\ \BBA {} Cohen%
}{%
Greene%
\ \protect \BOthers {.}}{%
{\protect \APACyear {2001}}%
}]{%
greene2001fmri}
\APACinsertmetastar {%
greene2001fmri}%
\begin{APACrefauthors}%
Greene, J\BPBI D.%
, Sommerville, R\BPBI B.%
, Nystrom, L\BPBI E.%
, Darley, J\BPBI M.%
\BCBL {}\ \BBA {} Cohen, J\BPBI D.%
\end{APACrefauthors}%
\unskip\
\newblock
\APACrefYearMonthDay{2001}{}{}.
\newblock
{\BBOQ}\APACrefatitle {An fMRI investigation of emotional engagement in moral
  judgment} {An fmri investigation of emotional engagement in moral
  judgment}.{\BBCQ}
\newblock
\APACjournalVolNumPages{Science}{293}{5537}{2105--2108}.
\PrintBackRefs{\CurrentBib}

\bibitem [\protect \citeauthoryear {%
Griffiths%
}{%
Griffiths%
}{%
{\protect \APACyear {2015}}%
}]{%
griffiths2015manifesto}
\APACinsertmetastar {%
griffiths2015manifesto}%
\begin{APACrefauthors}%
Griffiths, T\BPBI L.%
\end{APACrefauthors}%
\unskip\
\newblock
\APACrefYearMonthDay{2015}{}{}.
\newblock
{\BBOQ}\APACrefatitle {Manifesto for a new (computational) cognitive
  revolution} {Manifesto for a new (computational) cognitive
  revolution}.{\BBCQ}
\newblock
\APACjournalVolNumPages{Cognition}{135}{}{21--23}.
\PrintBackRefs{\CurrentBib}

\bibitem [\protect \citeauthoryear {%
Huang%
\ \protect \BOthers {.}}{%
Huang%
\ \protect \BOthers {.}}{%
{\protect \APACyear {2018}}%
}]{%
huang2018gpipe}
\APACinsertmetastar {%
huang2018gpipe}%
\begin{APACrefauthors}%
Huang, Y.%
, Cheng, Y.%
, Chen, D.%
, Lee, H.%
, Ngiam, J.%
, Le, Q\BPBI V.%
\BCBL {}\ \BBA {} Chen, Z.%
\end{APACrefauthors}%
\unskip\
\newblock
\APACrefYearMonthDay{2018}{}{}.
\newblock
{\BBOQ}\APACrefatitle {Gpipe: Efficient training of giant neural networks using
  pipeline parallelism} {Gpipe: Efficient training of giant neural networks
  using pipeline parallelism}.{\BBCQ}
\newblock
\APACjournalVolNumPages{arXiv preprint arXiv:1811.06965}{}{}{}.
\PrintBackRefs{\CurrentBib}

\bibitem [\protect \citeauthoryear {%
Khajah%
, Lindsey%
\BCBL {}\ \BBA {} Mozer%
}{%
Khajah%
\ \protect \BOthers {.}}{%
{\protect \APACyear {2016}}%
}]{%
khajah2016deep}
\APACinsertmetastar {%
khajah2016deep}%
\begin{APACrefauthors}%
Khajah, M.%
, Lindsey, R\BPBI V.%
\BCBL {}\ \BBA {} Mozer, M\BPBI C.%
\end{APACrefauthors}%
\unskip\
\newblock
\APACrefYearMonthDay{2016}{}{}.
\newblock
{\BBOQ}\APACrefatitle {How deep is knowledge tracing?} {How deep is knowledge
  tracing?}{\BBCQ}
\newblock
\APACjournalVolNumPages{arXiv preprint arXiv:1604.02416}{}{}{}.
\PrintBackRefs{\CurrentBib}

\bibitem [\protect \citeauthoryear {%
Krizhevsky%
, Sutskever%
\BCBL {}\ \BBA {} Hinton%
}{%
Krizhevsky%
\ \protect \BOthers {.}}{%
{\protect \APACyear {2012}}%
}]{%
krizhevsky2012imagenet}
\APACinsertmetastar {%
krizhevsky2012imagenet}%
\begin{APACrefauthors}%
Krizhevsky, A.%
, Sutskever, I.%
\BCBL {}\ \BBA {} Hinton, G\BPBI E.%
\end{APACrefauthors}%
\unskip\
\newblock
\APACrefYearMonthDay{2012}{}{}.
\newblock
{\BBOQ}\APACrefatitle {Imagenet classification with deep convolutional neural
  networks} {Imagenet classification with deep convolutional neural
  networks}.{\BBCQ}
\newblock
\BIn{} \APACrefbtitle {Advances in neural information processing systems}
  {Advances in neural information processing systems}\ (\BPGS\ 1097--1105).
\PrintBackRefs{\CurrentBib}

\bibitem [\protect \citeauthoryear {%
Linderman%
\ \BBA {} Gershman%
}{%
Linderman%
\ \BBA {} Gershman%
}{%
{\protect \APACyear {2017}}%
}]{%
linderman2017using}
\APACinsertmetastar {%
linderman2017using}%
\begin{APACrefauthors}%
Linderman, S\BPBI W.%
\BCBT {}\ \BBA {} Gershman, S\BPBI J.%
\end{APACrefauthors}%
\unskip\
\newblock
\APACrefYearMonthDay{2017}{}{}.
\newblock
{\BBOQ}\APACrefatitle {Using computational theory to constrain statistical
  models of neural data} {Using computational theory to constrain statistical
  models of neural data}.{\BBCQ}
\newblock
\APACjournalVolNumPages{Current opinion in neurobiology}{46}{}{14--24}.
\PrintBackRefs{\CurrentBib}

\bibitem [\protect \citeauthoryear {%
Luce%
}{%
Luce%
}{%
{\protect \APACyear {1959}}%
}]{%
luce1959individual}
\APACinsertmetastar {%
luce1959individual}%
\begin{APACrefauthors}%
Luce, R\BPBI D.%
\end{APACrefauthors}%
\unskip\
\newblock
\APACrefYear{1959}.
\newblock
\APACrefbtitle {Individual choice behavior: A theoretical analysis} {Individual
  choice behavior: A theoretical analysis}.
\PrintBackRefs{\CurrentBib}

\bibitem [\protect \citeauthoryear {%
Lázaro-Gredilla%
, Lin%
, Guntupalli%
\BCBL {}\ \BBA {} George%
}{%
Lázaro-Gredilla%
\ \protect \BOthers {.}}{%
{\protect \APACyear {2019}}%
}]{%
lazaro-gredilla_lin_guntupalli_george_2019}
\APACinsertmetastar {%
lazaro-gredilla_lin_guntupalli_george_2019}%
\begin{APACrefauthors}%
Lázaro-Gredilla, M.%
, Lin, D.%
, Guntupalli, J\BPBI S.%
\BCBL {}\ \BBA {} George, D.%
\end{APACrefauthors}%
\unskip\
\newblock
\APACrefYearMonthDay{2019}{}{}.
\newblock
{\BBOQ}\APACrefatitle {Beyond imitation: Zero-shot task transfer on robots by
  learning concepts as cognitive programs} {Beyond imitation: Zero-shot task
  transfer on robots by learning concepts as cognitive programs}.{\BBCQ}
\newblock
\APACjournalVolNumPages{Science Robotics}{4}{26}{}.
\PrintBackRefs{\CurrentBib}

\bibitem [\protect \citeauthoryear {%
McFadden%
\ \protect \BOthers {.}}{%
McFadden%
\ \protect \BOthers {.}}{%
{\protect \APACyear {1973}}%
}]{%
mcfadden1973conditional}
\APACinsertmetastar {%
mcfadden1973conditional}%
\begin{APACrefauthors}%
McFadden, D.%
\BCBT {}\ \BOthersPeriod {.}
\end{APACrefauthors}%
\unskip\
\newblock
\APACrefYearMonthDay{1973}{}{}.
\newblock
{\BBOQ}\APACrefatitle {Conditional logit analysis of qualitative choice
  behavior} {Conditional logit analysis of qualitative choice behavior}.{\BBCQ}
\newblock

\PrintBackRefs{\CurrentBib}

\bibitem [\protect \citeauthoryear {%
Mikhail%
}{%
Mikhail%
}{%
{\protect \APACyear {2002}}%
}]{%
mikhail2002aspects}
\APACinsertmetastar {%
mikhail2002aspects}%
\begin{APACrefauthors}%
Mikhail, J.%
\end{APACrefauthors}%
\unskip\
\newblock
\APACrefYearMonthDay{2002}{}{}.
\newblock
{\BBOQ}\APACrefatitle {Aspects of the theory of moral cognition: Investigating
  intuitive knowledge of the prohibition of intentional battery and the
  principle of double effect} {Aspects of the theory of moral cognition:
  Investigating intuitive knowledge of the prohibition of intentional battery
  and the principle of double effect}.{\BBCQ}
\newblock

\PrintBackRefs{\CurrentBib}

\bibitem [\protect \citeauthoryear {%
Mnih%
\ \protect \BOthers {.}}{%
Mnih%
\ \protect \BOthers {.}}{%
{\protect \APACyear {2015}}%
}]{%
mnih2015human}
\APACinsertmetastar {%
mnih2015human}%
\begin{APACrefauthors}%
Mnih, V.%
, Kavukcuoglu, K.%
, Silver, D.%
, Rusu, A\BPBI A.%
, Veness, J.%
, Bellemare, M\BPBI G.%
\BDBL {}others%
\end{APACrefauthors}%
\unskip\
\newblock
\APACrefYearMonthDay{2015}{}{}.
\newblock
{\BBOQ}\APACrefatitle {Human-level control through deep reinforcement learning}
  {Human-level control through deep reinforcement learning}.{\BBCQ}
\newblock
\APACjournalVolNumPages{Nature}{518}{7540}{529}.
\PrintBackRefs{\CurrentBib}

\bibitem [\protect \citeauthoryear {%
Nino%
}{%
Nino%
}{%
{\protect \APACyear {1983}}%
}]{%
nino1983consensual}
\APACinsertmetastar {%
nino1983consensual}%
\begin{APACrefauthors}%
Nino, C\BPBI S.%
\end{APACrefauthors}%
\unskip\
\newblock
\APACrefYearMonthDay{1983}{}{}.
\newblock
{\BBOQ}\APACrefatitle {A consensual theory of punishment} {A consensual theory
  of punishment}.{\BBCQ}
\newblock
\APACjournalVolNumPages{Philosophy \& Public Affairs}{}{}{289--306}.
\PrintBackRefs{\CurrentBib}

\bibitem [\protect \citeauthoryear {%
Quinn%
}{%
Quinn%
}{%
{\protect \APACyear {1989}}%
}]{%
quinn1989actions}
\APACinsertmetastar {%
quinn1989actions}%
\begin{APACrefauthors}%
Quinn, W\BPBI S.%
\end{APACrefauthors}%
\unskip\
\newblock
\APACrefYearMonthDay{1989}{}{}.
\newblock
{\BBOQ}\APACrefatitle {Actions, intentions, and consequences: The doctrine of
  doing and allowing} {Actions, intentions, and consequences: The doctrine of
  doing and allowing}.{\BBCQ}
\newblock
\APACjournalVolNumPages{The Philosophical Review}{98}{3}{287--312}.
\PrintBackRefs{\CurrentBib}

\bibitem [\protect \citeauthoryear {%
Rosenblatt%
}{%
Rosenblatt%
}{%
{\protect \APACyear {1958}}%
}]{%
rosenblatt1958perceptron}
\APACinsertmetastar {%
rosenblatt1958perceptron}%
\begin{APACrefauthors}%
Rosenblatt, F.%
\end{APACrefauthors}%
\unskip\
\newblock
\APACrefYearMonthDay{1958}{}{}.
\newblock
{\BBOQ}\APACrefatitle {The perceptron: a probabilistic model for information
  storage and organization in the brain.} {The perceptron: a probabilistic
  model for information storage and organization in the brain.}{\BBCQ}
\newblock
\APACjournalVolNumPages{Psychological review}{65}{6}{386}.
\PrintBackRefs{\CurrentBib}

\bibitem [\protect \citeauthoryear {%
Royzman%
\ \BBA {} Baron%
}{%
Royzman%
\ \BBA {} Baron%
}{%
{\protect \APACyear {2002}}%
}]{%
royzman2002preference}
\APACinsertmetastar {%
royzman2002preference}%
\begin{APACrefauthors}%
Royzman, E\BPBI B.%
\BCBT {}\ \BBA {} Baron, J.%
\end{APACrefauthors}%
\unskip\
\newblock
\APACrefYearMonthDay{2002}{}{}.
\newblock
{\BBOQ}\APACrefatitle {The preference for indirect harm} {The preference for
  indirect harm}.{\BBCQ}
\newblock
\APACjournalVolNumPages{Social Justice Research}{15}{2}{165--184}.
\PrintBackRefs{\CurrentBib}

\bibitem [\protect \citeauthoryear {%
Spranca%
, Minsk%
\BCBL {}\ \BBA {} Baron%
}{%
Spranca%
\ \protect \BOthers {.}}{%
{\protect \APACyear {1991}}%
}]{%
spranca1991omission}
\APACinsertmetastar {%
spranca1991omission}%
\begin{APACrefauthors}%
Spranca, M.%
, Minsk, E.%
\BCBL {}\ \BBA {} Baron, J.%
\end{APACrefauthors}%
\unskip\
\newblock
\APACrefYearMonthDay{1991}{}{}.
\newblock
{\BBOQ}\APACrefatitle {Omission and commission in judgment and choice}
  {Omission and commission in judgment and choice}.{\BBCQ}
\newblock
\APACjournalVolNumPages{Journal of experimental social
  psychology}{27}{1}{76--105}.
\PrintBackRefs{\CurrentBib}

\bibitem [\protect \citeauthoryear {%
Sutton%
\ \BBA {} Barto%
}{%
Sutton%
\ \BBA {} Barto%
}{%
{\protect \APACyear {1981}}%
}]{%
sutton1981toward}
\APACinsertmetastar {%
sutton1981toward}%
\begin{APACrefauthors}%
Sutton, R\BPBI S.%
\BCBT {}\ \BBA {} Barto, A\BPBI G.%
\end{APACrefauthors}%
\unskip\
\newblock
\APACrefYearMonthDay{1981}{}{}.
\newblock
{\BBOQ}\APACrefatitle {Toward a modern theory of adaptive networks: expectation
  and prediction.} {Toward a modern theory of adaptive networks: expectation
  and prediction.}{\BBCQ}
\newblock
\APACjournalVolNumPages{Psychological review}{88}{2}{135}.
\PrintBackRefs{\CurrentBib}

\bibitem [\protect \citeauthoryear {%
Thomson%
}{%
Thomson%
}{%
{\protect \APACyear {1984}}%
}]{%
thomson1984trolley}
\APACinsertmetastar {%
thomson1984trolley}%
\begin{APACrefauthors}%
Thomson, J\BPBI J.%
\end{APACrefauthors}%
\unskip\
\newblock
\APACrefYearMonthDay{1984}{}{}.
\newblock
{\BBOQ}\APACrefatitle {The trolley problem} {The trolley problem}.{\BBCQ}
\newblock
\APACjournalVolNumPages{Yale Law Journal}{94}{}{1395}.
\PrintBackRefs{\CurrentBib}

\bibitem [\protect \citeauthoryear {%
Tversky%
\ \BBA {} Simonson%
}{%
Tversky%
\ \BBA {} Simonson%
}{%
{\protect \APACyear {1993}}%
}]{%
tversky1993context}
\APACinsertmetastar {%
tversky1993context}%
\begin{APACrefauthors}%
Tversky, A.%
\BCBT {}\ \BBA {} Simonson, I.%
\end{APACrefauthors}%
\unskip\
\newblock
\APACrefYearMonthDay{1993}{}{}.
\newblock
{\BBOQ}\APACrefatitle {Context-dependent preferences} {Context-dependent
  preferences}.{\BBCQ}
\newblock
\APACjournalVolNumPages{Management science}{39}{10}{1179--1189}.
\PrintBackRefs{\CurrentBib}

\bibitem [\protect \citeauthoryear {%
Woollard%
\ \BBA {} Howard-Snyder%
}{%
Woollard%
\ \BBA {} Howard-Snyder%
}{%
{\protect \APACyear {2016}}%
}]{%
sep-doing-allowing}
\APACinsertmetastar {%
sep-doing-allowing}%
\begin{APACrefauthors}%
Woollard, F.%
\BCBT {}\ \BBA {} Howard-Snyder, F.%
\end{APACrefauthors}%
\unskip\
\newblock
\APACrefYearMonthDay{2016}{}{}.
\newblock
{\BBOQ}\APACrefatitle {Doing vs. Allowing Harm} {Doing vs. allowing
  harm}.{\BBCQ}
\newblock
\BIn{} E\BPBI N.~Zalta\ (\BED), \APACrefbtitle {The Stanford Encyclopedia of
  Philosophy} {The stanford encyclopedia of philosophy}\ (\PrintOrdinal{Winter
  2016}\ \BEd).
\newblock
\APACaddressPublisher{}{Metaphysics Research Lab, Stanford University}.
\PrintBackRefs{\CurrentBib}

\bibitem [\protect \citeauthoryear {%
Yarkoni%
\ \BBA {} Westfall%
}{%
Yarkoni%
\ \BBA {} Westfall%
}{%
{\protect \APACyear {2017}}%
}]{%
yarkoni2017choosing}
\APACinsertmetastar {%
yarkoni2017choosing}%
\begin{APACrefauthors}%
Yarkoni, T.%
\BCBT {}\ \BBA {} Westfall, J.%
\end{APACrefauthors}%
\unskip\
\newblock
\APACrefYearMonthDay{2017}{}{}.
\newblock
{\BBOQ}\APACrefatitle {Choosing prediction over explanation in psychology:
  Lessons from machine learning} {Choosing prediction over explanation in
  psychology: Lessons from machine learning}.{\BBCQ}
\newblock
\APACjournalVolNumPages{Perspectives on Psychological
  Science}{12}{6}{1100--1122}.
\PrintBackRefs{\CurrentBib}

\end{thebibliography}

\end{document}